\documentclass[twocolumn,prl]{revtex4}

\usepackage{amsmath,amsfonts}

\newcommand{\nuc}[2]{$^{#1}$#2}

\begin{document}

\title{Computer-generated character tables and nuclear spin
  statistical weights: Application to benzene dimer and methane
  dimer}

\author{Roman~Schmied}
\author{Kevin~K.~Lehmann}
\affiliation{Department of Chemistry, Princeton University, Princeton,
  NJ 08544}

\date{\today}

\maketitle

In the assignment of high-resolution spectra of van der Waals
molecular clusters, it is of great utility to know the character table
of the molecular symmetry group and the nuclear spin weights of the
various rovibronic symmetry species~\cite{Bunker1998}. While their
calculation is not difficult, for large groups it is
lengthy~\cite{Balasubramanian1991} and prone to errors due to the
sheer quantity of variables involved.  For the benzene dimer, both the
character table~\cite{Odutola1981} and the nuclear spin
weights~\cite{Spirko1999} have been published with errors (though the
latter had previously been published
correctly~\cite{Balasubramanian1984}); for methane dimer, the nuclear
spin weights have been published with an
error~\cite{Balasubramanian1991}.

We would like to bring to the attention of the spectroscopy community
a free software package for group theory named GAP~\cite{GAP4}, which
greatly facilitates these calculations. As an example of its usage, we
present calculations for the permutation-inversion (PI) groups of
benzene dimer, which is the direct product of a permutation (P) group
and the inversion group $\{E,E^*\}$, and of methane dimer, for which a
planar structure is not accessible and thus $E^*$ is not a feasible
operation.  For both dimers, rigid monomer units are assumed.
Calculations for methane and benzene trimers, and water hexamer, can
be done in this way within a few seconds on a personal computer.

In what follows, we consider only the hydrogen atoms, since the
\nuc{12}{C} isotope has nuclear spin zero. We label the hydrogen atoms
on the two benzene monomers $1\dotsc6$ and $7\dotsc12$, in a circular
way.  The permutation group is specified by its generators, for
example $C_6$, $C_2'$, and $\tau$ (monomer exchange), by entering at
the GAP prompt:
\begin{verbatim}
  b2P := Group( (1,2,3,4,5,6), (2,6)(3,5),
         (1,7)(2,8)(3,9)(4,10)(5,11)(6,12) );
\end{verbatim}
In the case of methane dimer, we introduce two fictitious atoms, 9 and
10, with nonphysical negative spin quantum numbers, whose interchange
symbolizes the inversion. Generators of the PI group are $C_3$,
$\sigma^*$ (reflection-inversion), $C_2$, and $\tau$:
\begin{verbatim}
  m2PI := Group( (1,2,3), (1,2)(5,6)(9,10),
          (1,2)(3,4), (1,5)(2,6)(3,7)(4,8) );
\end{verbatim}
The character table for benzene dimer is requested with
\verb|Display(CharacterTable(b2P))|, and the conjugacy classes with
\verb|ConjugacyClasses(CharacterTable(b2P))|.  Mind however that the
resulting sorting of the conjugacy classes and irreducible
representations is different from that of Ref.~\cite{Odutola1981}.

The computation of the nuclear spin weights of the various rovibronic
symmetry species is computed using the formula of
Ref.~\cite{Jonas1989}. For each permutation cycle of length $n$,
permuting atoms of spin $i$, there is a factor of $(2i+1)
(-1)^{2i(n-1)}$, except if the cycle symbolizes an inversion, in which
case the character value is zero:
\begin{verbatim}
  CycleFactor := function(n,i)
     if i < 0 then return 2-n;
        else return (2*i+1)*(-1)^(2*i*(n-1));
     fi;  end;
\end{verbatim}
The rovibronic character value of a permutation $p$ acting on a list
of atoms $k$ with spins $s$ is the product of the above factors for
each cycle in the permutation, multiplied by 2 (for the two parity
labels):
\begin{verbatim}
  rveCharacterVal := function(p,k,s)
     return 2*Product(Cycles(p,k),
        c -> CycleFactor(Length(c),s[c[1]]));
  end;
\end{verbatim}
Looping this function over the conjugacy classes of a group $g$ yields
the character of the allowed rovibronic wave functions:
\begin{verbatim}
  rveCharacter := function(g,s)
     return List(ConjugacyClasses(g),
        p -> rveCharacterVal(Elements(p)[1],
             MovedPoints(g),s));  end;
\end{verbatim}
Finally, decomposition of this character into irreducible characters
yields the nuclear spin weights of the different rovibronic species:
\begin{verbatim}
  rveWeights := function(g,s)
     return MatScalarProducts(Irr(g),
        [rveCharacter(g,s)])[1];  end;
\end{verbatim}

The spin weights of (C$_6$H$_6$)$_2$ are computed with
\verb|rveWeights(b2P,0*[1..12]+1/2)|, and those of (C$_6$D$_6$)$_2$
with \verb|rveWeights(b2P,0*[1..12]+1)|. Since this calculation is
based on the P group, parity labels must be added to each symmetry
species displayed in the character table, and the above spin weights
are evenly split between even and odd symmetry species. The character
table of benzene dimer has been published previously with errors in
Ref.~\cite{Odutola1981}: the signs of the character values of
$C_3C_3$, line 10; $C_6C_6$, line 12; and $C_2''C_3$, line 24, need to
be inverted. In the ordering of that table, the rovibronic spin
weights of (C$_6$H$_6$)$_2$ are 28, 21, 6, 3, 78, 91, 1, 0, 21, 91, 7,
39, 3, 13, 66, 55, 45, 36, 77, 63, 33, 27, 143, 117, 11, 9, 99, and
those of (C$_6$D$_6$)$_2$ are 4278, 4186, 741, 703, 2628, 2701, 1081,
1035, 3496, 6716, 4232, 2774, 1748, 3358, 6786, 6670, 7750, 7626,
10672, 11408, 4408, 4712, 8468, 9052, 5336, 5704, 14384.

The spin weights of (CD$_4$)$_2$ are computed with
\verb|rveWeights(m2PI,[1,1,1,1,1,1,1,1,-1,-1])|, where the last two
spins are unphysical and designate the fictitious atoms whose exchange
stands for the inversion operation; the same calculation with spins
$1/2$ computes the spin weights of (CH$_4$)$_2$. The character table
of methane dimer has been published previously~\cite{Odutola1981}; in
that ordering, the rovibronic spin weights of (CH$_4$)$_2$ are 15, 15,
10, 10, 2, 0, 1, 1, 10, 15, 15, 6, 6, 3, 3, 6, and those of
(CD$_4$)$_2$ are 120, 120, 105, 105, 42, 30, 36, 36, 180, 270, 270,
171, 171, 153, 153, 216.

This work was supported by the National Science Foundation.

\end{document}